\documentclass[twocolumn,aps,pra]{revtex4}
\usepackage{epsfig}
\usepackage{psfig}

\begin{document}

\title{Pseudoresonance mechanism of all-optical frequency standard operation}
\author{G.Kazakov$^{1}$\thanks{E--mail:
kazakov@quark.stu.neva.ru}, B.Matisov$^{1}$, I.Mazets$^{2}$,
G.Mileti$^{3}$, J.Delporte$^{4}$}
\affiliation{{\setlength{\baselineskip}{18pt}
$^{1}${St. Petersburg State Polytechnic University,
St. Petersburg, 195251, Russia}\\
$^{2}${A.F. Ioffe Physics-Technical Institute,
St. Petersburg 194021, Russia}\\
$^3${Observatoire Cantonal Neuch\^{a}tel,
2000 Neuch\^{a}tel, Switzerland}\\
$^4${Centre National d'Etudes Spatiales, 31401  Toulouse, France}
}}

\begin{abstract}
We propose a novel approach to all-optical frequency standard
design, based on a counterintuitive combination of the coherent
population trapping effect and signal discrimination at the
maximum of absorption for the probe radiation. The short-term
stability of such a standard can achieve the level of ${10^{ -14}
/ \sqrt {\tau}}$. The physics beyond this approach is dark resonance
splitting caused by interaction of the nuclear magnetic moment with
the external magnetic field.
\\
\vspace*{10pt}
\end{abstract}
\maketitle

The unit of time in the Syst\`eme International (the second) is
defined via the period of the transition between the hyperfine
(HF) components of the ground state of the $^{133}$Cs atom (in the
limit of vanishing external perturbations). Secondary standards
may use other elements (alkali metals, hydrogen). Operation of a
frequency standard is provided by locking a microwave signal produced by a
quartz crystal oscillator to
a resonance on the transition between the
working levels \cite{x14}, i. e. the Zeeman components of the HF structure
with zero projection of the total angular momentum. Such a choice
of the working levels is dictated by the absence of the linear
Zeeman shift. The resonance can be excited by direct microwave
transition or by two-photon Raman transition. The latter option
provides the basis for all-optical frequency standards
\cite{y4,y1,y2,y3,x4,x11,x1}
where the setup includes no microwave cavities,
but electro-optical modulation of the laser beam or
direct modulation of the diode laser current are used instead

If the interaction time of an atom with the laser field exceeds
few optical pumping cycles then the coherent population trapping
(CPT) sets in, giving rise to the so-called dark resonance
\cite{x2,x3}.
The physical reason for the CPT is optical pumping of atoms into a
coherent superposition of the two ground state sublevels, which is
immune to excitation by the frequency-split laser radiation. The
CPT leads to decrease of absorption of the laser light. However,
if the Raman detuning (the difference between the frequency
splitting of the two-component laser radiation and the transition
frequency between the two ground state sublevels) exceeds the dark
resonance width (what can be as small as few dozens Hz
\cite{x4,x5}), the
CPT atomic state is destroyed, and usual value of the laser
radiation absorption is restored.

Keeping the frequency of a generator that provides
laser frequency splitting
coincident with the position of the dark resonance on the working transition
is the physical mechanism of the frequency standardization by optical means.

In the present paper we propose a different, quite
counterintuitive technique. No specific resonance on the working
transition is excited. Instead, its frequency is identified as
\textit{the position of maximum absorption between the two side
dark resonances involving magnetically-sensitive Raman
transitions}.

Before to proceed further, we have to recall some basic ideas from
the theory of CPT in systems with a closed-loop interaction
contour \cite{x6}. An example of such a system is given in Fig. 1.

The reason why $\sigma ^{ +}  / \sigma ^{ -} $-configuration of the laser
radiation polarizations is used, instead of simply applying
$\sigma ^{+}$-polarized light only,
is the adverse influence of the states with the
maximum projection of the angular momentum, where most of the atoms are
accumulated, this drastically reducing the dark resonance contrast and,
hence, worsening the standard's performance.

Consider the pair of the ground (g) state sublevels, ${| {F_{g} =
F,\,\,\,m_{g} = m} \rangle} $ and ${| {F_{g} = F + 1,\,\,\,m_{g} =
m} \rangle} $, each of them resonantly coupled to the first
excited (e) state sublevels ${| {F_{e} = F,\,\,\,m_{e} = m + 1}
\rangle }$ and ${| {F_{e} = F,\,\,\,m_{e} = m - 1} \rangle} $ (or
${| {F_{e} = F + 1,\,\,\,m_{e} = m + 1} \rangle}$ and $| F_{e} = F
+ 1,\,\,m_{e} = m - 1 \rangle $), this forming so-called double $
\Lambda $ closed loop interaction contour. Here $F$ is the total
angular momentum of the given HF component, $m$ is the
angular momentum
projection to the quantization axis. The interaction matrix
element for the transition ${| {F_{g},\,m_{g}} \rangle}
\leftrightarrow {| {F_{e} ,\,m_{e}} \rangle} $ is
\begin{equation}
 \langle F_{g},\,m_{g} |
\hat{V} |F_{e} ,\,m_{e}  \rangle  = - \langle F_{g} ,\,m_{g} |
\hat{d} _{q}  |F_{e} ,\,m_{e}  \rangle E_{F_{g},\, F_{e},\, q} ,
\label{eq1}
\end{equation}
where $E_{F_{g} ,\,F_{e}, \,q} $ is the complex amplitude of
the laser field resonant to the given transition, $q = m_{e} -
m_{g} $ denotes the cyclic component of the polarization unit
vector, and the electric dipole element can be represented,
according to the Wigner-Eckart theorem \cite{x7} as
\begin{eqnarray}
\langle F_{g} ,m_{g} |
\hat{d} _q|F_{e} ,m_{e}  \rangle = ( - 1)^{F_{g} + J_{e} + I -
1}\sqrt{2F_{g} + 1} && \nonumber \\
\times  C_{F_{g} \,m_{g}
\,\,1\,q}^{F_{e} \,m_{e}}
\left \{
\begin{array}{ccc}
J_{g}   & I  & F_{g}   \\
F_{e}   & 1  & J_{e}
\end{array}
\right \}
\langle J_{g} ||\hat{d} ||J_{e} \rangle ,&&
\label{eq2}
\end{eqnarray}
where $J_{g(e)} $ is the electronic
angular momentum of the ground (excited) state,
$I$ is the nuclear spin, $C_{F_{g} \,m_{g} \,\,1\,q}^{F_{e} \,m_{e}}  $ is
the Clebsh-Gordan coefficient, $\left \{ \begin{array}{ccc}
J_{g}   & I  & F_{g}   \\
F_{e}   & 1  & J_{e}   \end{array} \right \} $
is the $6j$-symbol, and $\langle J_{g} ||
\hat{d} ||J_{e}  \rangle $ is the reduced matrix element of the
dipole moment.

Consider the two $\Lambda $-schemes produced by the laser radiation with
$\sigma ^{ +} $ and $\sigma ^{ -} $ polarizations separately. The
two corresponding CPT states, $| CPT_+ \rangle $
and $|CPT _- \rangle $, can be identified as
\begin{eqnarray}
\lefteqn{ | CPT_{\pm } \rangle = \left( \langle F_{g} = F
+ 1,m|\hat{V}|F_{e} ,m \pm 1 \rangle |F_{g}=F,m\rangle-\right.}
&& \nonumber \\
&&\left. -  \langle F_{g} = F ,m|\hat {V}|F_{e} ,m \pm 1 \rangle
|F_{g}=F+1,m\rangle\right)\times \nonumber \\
&&\times \left( |\langle F_{g} = F ,m|\hat{V}|F_{e} ,m \pm 1
\rangle|^{2}+\right. \nonumber \\
&&\left. +|\langle F_{g} = F + 1,m|\hat{V}|F_{e} ,m \pm 1
\rangle|^{2} \right)^{-1/2}.
\label{eq3a}
\end{eqnarray}

Let $\varsigma $ be a parameter defined via the relation
\begin{eqnarray}
\label{eq5a}
\lefteqn{ {\frac{{\left\langle {F_{g} = F + 1,m{\left|
\hat{V} \right|}F_{e} ,m + 1} \right\rangle}} {\left\langle {F_{g}
= F,m{\left|\hat{V}\right|}F_{e} ,m + 1} \right\rangle}} =\qquad } &&
\nonumber \\
&& =\varsigma
{\frac{{\left\langle {F_{g} = F + 1,m{\left| \hat{V}
\right|}F_{e} ,m - 1} \right\rangle}} {\left\langle {F_{g} =
F,m{\left|\hat{V}\right|}F_{e} ,m - 1} \right\rangle}} .
\end{eqnarray}
If the amplitude and phase conditions \cite{x6,x8} for the complex
interaction matrix elements are satisfied, i. e.  $\varsigma =+1$,
then the two states defined by Eq. (\ref{eq3a}) coincide,
and we get perfect CPT in the double $\Lambda $-scheme. Then, if
we scan the Raman detuning across the resonance, the dark
resonance in the absorption spectrum appears \cite{x9}. In the
contrary, if $\varsigma =-1$
then ${\left| {CPT_{ +}}   \right\rangle} $ is the state corresponding to
maximally enhanced absorption of the $\sigma ^{ -} $ -polarized radiation,
and \textit{vice versa}.
It makes CPT in such a double $\Lambda $-scheme impossible. A simple
but a bit lengthy analysis shows that no structure appears in the absorption
spectrum near the resonance, \textit{if no other ground state
sublevels are taken into account}.

Unfortunately, satisfying the condition $\varsigma =+1$ implies
sophisticated experimental techniques. Indeed, if all the four
components of the laser radiation, different in polarization and
frequency, are obtained from the same input light beam by
electro-optical modulator without specific precautions, one gets
${{E_{F,\,F_{e} ,\, + 1}}  \mathord{\left/ {\vphantom
{{E_{F,\,F_{e} ,\, + 1}}  {E_{F + 1,\,F_{e} ,\, + 1}}} } \right.
\kern-\nulldelimiterspace} {E_{F + 1,\,F_{e} ,\, + 1}}}  =
{{E_{F,\,F_{e} ,\, - 1}}  \mathord{\left/ {\vphantom
{{E_{F,\,F_{e} ,\, - 1}}  {E_{F + 1,\,F_{e} ,\, - 1}}} } \right.
\kern-\nulldelimiterspace} {E_{F + 1,\,F_{e} ,\, - 1}}} $. Due to
the general property of Clebsh-Gordan coefficients \cite{x7}, we always
get for the working transition
\begin{eqnarray}
\lefteqn{ \langle F,\,0| \hat{d} |F_{e}, + 1 \rangle \left/\langle
F+1,\,0 | \hat{d} |F_{e}, +
1 \rangle=\right.} \quad && \nonumber \\
&& = -\langle F,\,0| \hat{d} |F_{e}, - 1 \rangle \left/\langle
F+1,\,0 | \hat{d} |F_{e}, - 1\rangle \right.
\label{eqnew1}
\end{eqnarray}
that results in $\varsigma =-1$. Therefore one has to
introduce a $\pi $ phase shift to the $\sigma ^{ -} $-component of
the laser radiation, with respect to $\sigma ^{ +} $ one. One
solution, implemented by the NIST group
\cite{x10,x11}, is based on
counterpropagating configuration of
$\sigma ^{ +} $- and $\sigma ^{ -} $-polarized
beams. In this case, the CPT condition $\varsigma =+1$
is satisfied in periodically located
spatial regions, which are quite narrow (much less than a quarter
of the HF transition wavelength). It limits the cell size by $\sim
$ 1 mm. Dark resonance in such a miniaturized cell are quite
broad, and the corresponding short-term stability of a frequency
standard is limited by $\sim {10^{ - 11}} / {\sqrt {\tau}}$
\cite{y1,x11}, $\tau $ being the
integration time (in seconds). Recently high-contrast dark
resonance have been demonstrated using so-called ``push-pull''
technique \cite{x12}, where a set of $\lambda / 4$ plates,
polarization-sensitive beam splitters and reflectors provides the
necessary phase shift in the cw regime. The ``lin $ \bot $ lin''
technique developed by Zanon et. al. \cite{x13} realizes a similar idea
in time-domain (Ramsey) spectroscopy. The setups of Refs. \cite{x12,x13}
are quite good for high-precision \textit{laboratory} frequency
standards. However, it is quite desirable to design a simpler in
construction and easier to handle all-optical frequency standard,
more suitable, e.g., for operation on board of a satellite or in
other mobile instrument applications.
In the present paper we propose a relevant approach.

\begin{figure}
\begin{center}
\psfig{file=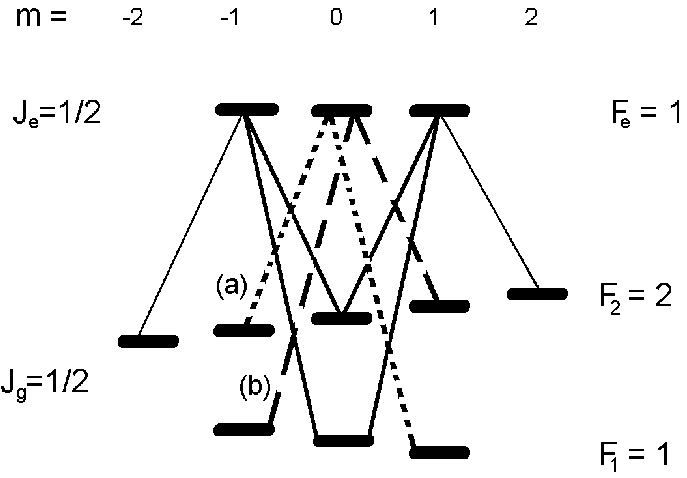,width=7.5cm}
\end{center}
\begin{caption}
{Scheme of optically-induced transitions in $^{87}$Rb
atom, case of $F_{e} = 1$. ``Closed loops'' involving the working
(0-0) transition are shown by bold solid black lines. Additional
$\Lambda $-schemes are shown by (a) dotted lines (the pair
of ground state Zeeman sublevels $\vert F_{g}$ = 1, $m$ =
+1$\rangle $, $\vert F_{g}$ = 2, $m$ = --1$\rangle $
being involved), and (b)
dashed lines ($\vert F_{g}$ = 1, $m$ = --1$\rangle $,
$\vert F_{g}$ = 2, $m$ = +1$\rangle $ being involved).}
\end{caption}
\label{f:1}
\end{figure}

We have obtained an intriguing result: the ``adverse''
condition $\varsigma =-1$ itself,
holding for $m = 0$, does not prevent precise spectroscopic
identification of the working transition frequency and locking to
it a quartz generator. The key factor here is the influence of the
other Zeeman sublevels of the HF structure of the ground state.
For the sake of definiteness, let us consider gas of $^{87}$Rb
atoms placed in a homogeneous magnetic field \textbf{H}. A
linearly polarized laser radiation propagates along \textbf{H}.
Its carrier frequency and modulation frequency are chosen so that
the $F_{g} = 1 \leftrightarrow F_{e} = 1$ and $F_{g} = 2
\leftrightarrow F_{e} = 1$ components of the D$_{1}$-line are
excited. Being projected to the quantization axis defined by the
magnetic field, the linear polarization is represented by the
coherent superposition of $\sigma ^{ +} $ and $\sigma ^{ -}
$-polarizations, see Fig. 1. Since there is no delay line for the
$\sigma ^{ +} $-polarized component, we have $\varsigma =-1$ for $m
= 0$, and the corresponding double $\Lambda $-scheme (shown by bold
lines in Fig.~1) exhibit no
CPT resonance. However, there are two additional $\Lambda
$-schemes, corresponding to quadrupole ($\left| \Delta m \right| =2$)
two-photon transitions and
involving the pairs of Zeeman sublevels {\{}$\vert
F_{g}$ = 1, $m$ = +1$\rangle $, $\vert F_{g}$ = 2, $m$ =
--1$\rangle ${\}}, indicated by (a) Fig. 1, and {\{}$\vert
F_{g}$ = 1, $m$ = --1$\rangle $, $\vert F_{g}$ = 2, $m$ =
+1$\rangle ${\}}, indicated by (b) in Fig. 1.
The distances
between these ground state Zeeman sublevels in the weak magnetic
field $H$ are \cite{x14}
\begin{equation}
\omega _{a,b}  =
\omega _{hfs} \pm {\frac{{2g_{I} \mu _{N} }}{{\hbar}} }H +
{\frac{{3g_{J}^{2} \mu _{B}^{2}}} {{\,8\omega _{hfs} \,\hbar
^{2}}}}H^{2},
\label{eq7}
\end{equation}
where \textit{$\omega $}$_{hfs}$ is the hyperfine splitting
frequency of the ground state HF components in the absence of
magnetic field, $\mu _B=e\hbar $/(2$m_{e}c)$ is the Bohr magneton,
\textit{$\mu $}$_{N}=e\hbar $/(2$m_{p} c)$ is the nuclear magneton,
$g_{I}=\mu / \mu _{N}$ is the nuclear Lande factor, $g_{J}$ is the
electronic Lande factor, $\mu $ is the magnetic momentum of atomic
nucleus. The positive sign of the second term in the right-hand-side
of Eq.~(\ref{eq7}) corresponds to the $\Lambda $-scheme denoted in
Fig. 1 by (a), the negative sign corresponds to the scheme (b). Note
that one may erroneously treat these side dark resonances as
magnetically insensitive in the linear approximation, if one does
not take into account the magnetic moment of the nucleus, since the
linear Zeeman shifts due to the electronic magnetic moment for each
state in the pair $\vert F_{g}$ = 1, $m=\pm $1$\rangle $, $\vert
F_{g}$ = 2, $m=\mp $1$\rangle $ are the same. Observation of
splitting of dark resonances due to interaction of nuclear spins
with external magnetic field was first reported in Ref. \cite{csns}
for Cs atoms, and recently \cite{vy} for $^{87}$Rb atoms. In Eq.
(\ref{eq7}) we retain, for the sake of generality, the second-order
Zeeman effect terms.

Therefore, magnetic field shifts the positions of the corresponding dark
resonances close to the working transition. The two dark resonances are
symmetric (to the second order Zeeman effect) with respect to the position
of the ${\left| {F_{g} = 1,m = 0} \right\rangle}  \leftrightarrow {\left|
{F_{g} = 2,m = 0} \right\rangle} $ working transition. If the magnetic field
shift is small (in comparison with single resonance width), the two dark
resonances are not resolved, see Fig. 2,
the dotted curve. For intermediate values
of $H$ they are resolved partially, thus forming a minimum in the
transmission spectrum exactly between them, see Fig. 2,
the solid curve. The
position of this minimum coincides with the frequency of the ${\left|
{F_{g} = 1,m = 0} \right\rangle}  \leftrightarrow {\left| {F_{g} = 2,m = 0}
\right\rangle} $ transition and can be implemented in discriminator of an
all-optical frequency standard. Such a structure formed between two
maxima in the transmission spectrum
can be called a \textit{pseudoresonance}.
The minimum width of the pseudoresonance
is of the order of the dark resonance width. Further increase of ${H}$
makes the two dark resonances fully resolved. As a result, the bottom
of the pseudoresonance dip becomes flat, which is unsuitable for the
frequency standardization purposes, see Fig.2, the dot-dashed curve.
The second-order Zeeman
shift of the pseudoresonance is of order of the same shift for the working
transition.

\begin{figure}
\begin{center}
\psfig{file=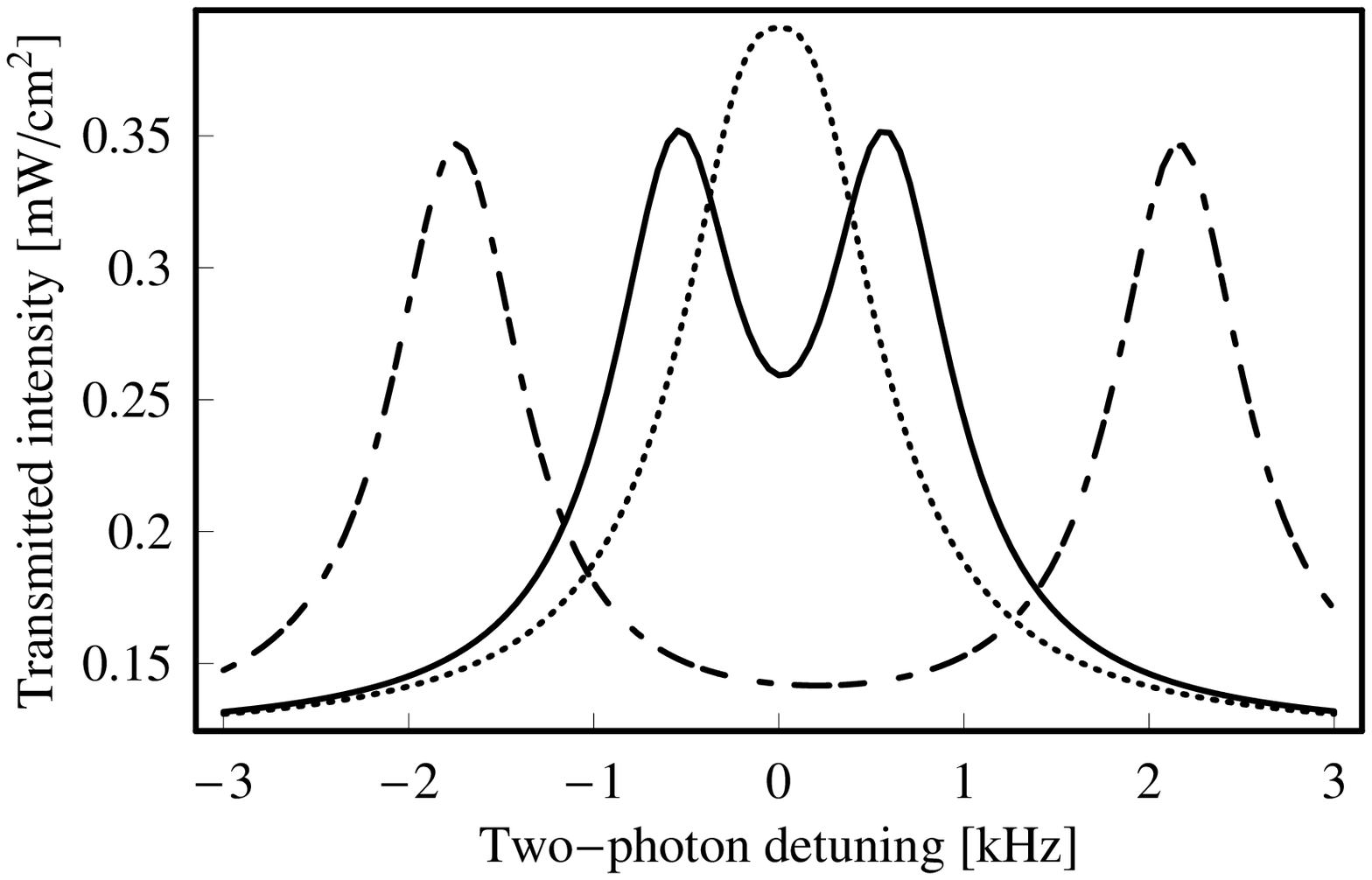, width=7.8cm}
\end{center}
\begin{caption}{Transmission spectrum of the cell with the parameters
specified in the text.
Incident radiation intensity intensity $U_0 = 0.5$ mW/cm$^{2}$.
The excitation scheme is shown in Fig.~1.
Dotted line corresponds to $H$ = 0.05 G, solid
line  corresponds to $H$ = 0.2 G, dot-dashed line  corresponds
to $H$ = 0.7 G. }\end{caption}
\label{f:2}
\end{figure}

\begin{figure}
\begin{center}
\psfig{file=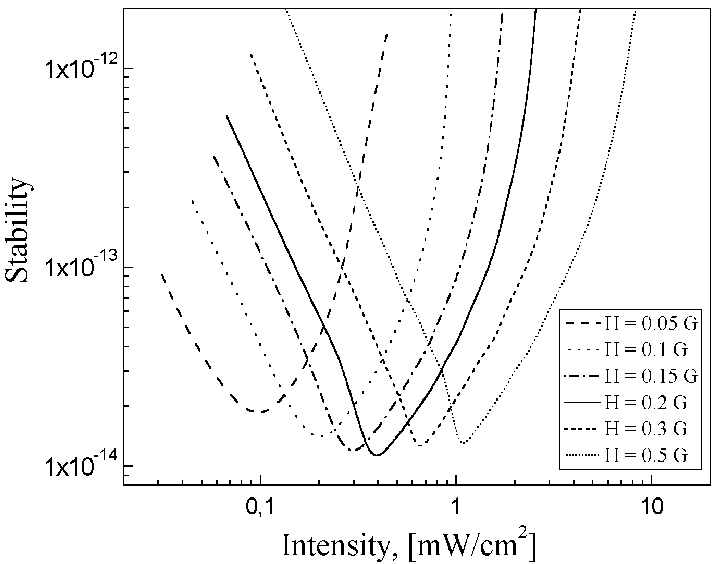, width=7.8cm}
\end{center}
\begin{caption}{Frequency standard stability
(Allan deviation, dimensionless) for the
integration time $\tau = 1\,\,s$ versus laser field intensity for
different magnetic fields. }\end{caption}
\label{f:3}
\end{figure}

The light shift  of the pseudoresonance position is the same as
that of each of the two side dark resonances \cite{lsh1,lsh2}. In
particular, it vanishes if the laser is tuned exactly in resonance
with the $F_e=1$ component of the $^{87}$Rb D$_1$-line.

Note that the $F_e=2$ component of the $^{87}$Rb
D$_1$ line can not be used for
pseudoresonance creation, since in this case the states
$\left| F_e=2, \, m=\pm 2\right \rangle $ are excited thus
turning the relevant $\Lambda $-schemes into $W$-type schemes, in which
CPT is not possible \cite{x2}. D$_2$-line is also unsuitable because of
the small magnitude of the HF splitting of the
$5\, ^2P^\circ _{3/2}$ excited state
and, hence, strong overlapping
of Doppler contours of optical transitions to different HF components
of the excited state.

We performed numerical simulations of linearly-polarized,
two-component laser light propagation through a $^{87}$Rb gas cell
with the following parameters: radius $R_c=1$ cm, length
$L_c=2.5$ cm, number
density $n = 1.1 \times 10^{11}$ cm$^{-3}$ ($T = 327$ K), the diffusion
coefficient for $^{87}$Rb atoms $D =
20$ cm$^{2}$/s corresponding to the buffer gas (N$_2$)
pressure = 15 Torr (such a pressure is chosen to provide
quenching of the excited state and thus diminish rescattering of
photons). The
relaxation rate $\Gamma $ of ground state coherences in such a cell is
dominated by wall collisions \cite{x14} and can be estimated as
$\Gamma \approx 300$~s$^{-1}$. In the present paper we do not discuss
the effect of pressure shift of the HF transition \cite{x14,x15} and
the choice of an optimum buffer gas mixture, postponing this subject to
a more extended publication. However, we aware that one has to
minimize the thermal dependence of the pressure shift,
especially for improving the long-term stability of the standard.
As concerns Fig. 2, the pressure shift at the specified cell temperature
is included there into the working transition frequency value for
zero magnetic field.
The density matrix formalism for the whole manifold of the states was
applied. The results are shown in Fig. 2. Apparently, the width of
the absorption peak of about 0.7 kHz is achievable simultaneously
with the contrast $\sim  30$ {\%}. In Fig. 3 we plot the stability
$\sigma _{y} $ of the standard based on the
\textit{pseudoresonance} discrimination method versus the incident
radiation intensity for different values of external magnetic
fields \cite{x14}:
\begin{equation}
\sigma _{y} \approx \left( {\omega _{hfs}} \left| S^{\prime \prime }
(0)\right| Wf(\Theta )\sqrt
{{\frac{{P\tau}} {{\hbar \omega _{0}}} }}  \right)^{ - 1},
\label{eqnew2}
\end{equation}
where ${P = U_0A}$ is the total laser
radiation power, $U_0$ being the intensity at the
cell input, and $A$ being the
effective beam cross-section area (in our simulations we assume
that the laser beam is wide enough to provide $A\approx \pi R_c^2$),
\textit{$\tau $} is the integration time, $\omega _{0}$ is the
D$_{1}$-line resonance frequency, $S^{\prime \prime }(0)$ is the
second derivative of the signal over the two-photon (Raman) detuning
at the extremum of the reference spectroscopic line
(in our case, at the pseudoresonance position), $W$ is the width of the
detuning range where $S^{\prime \prime }$ is approximately constant.
The signal $S$ is the transmitted laser power normalized to its value
outside the CPT-resonance induced structure but still within the
Doppler profile of the D$_1$-line. The function $f(\Theta )$ is close to
1 in the optimal range of values of the cell optical thickness $\Theta $
($0.5<\Theta <1$). Standard
stability estimation shows that short-term stability $\sigma _{y} \sim
10^{-14} /{\sqrt {\tau}} $ is achievable.

To conclude, we demonstrate theoretically the existence of a novel
structure in a laser-light transmission spectrum under
the CPT condition, a \textit{pseudoresonance}, that is a narrow
(having the width much less than the natural width of the
excited state)
absorption maximum between the two dark resonances involving $m =
\pm 1$ sublevels of the ground state of an alkali atom, while CPT
in the double $\Lambda $-scheme based on $m = 0$ is absent. The
performance of an all-optical frequency standard using this effect
for signal discrimination must be the same, as that of a standard
employing the ``push-pull'' technique \cite{x12}. However, a
pseudoresonance-based standard is to be easier in both construction and
handling and thus more suitable for application on board of a
satellite or in other mobile instrument applications.

This research is supported by the INTAS-CNES, project 03-53-5175,
and by the Ministry of Education and Science of Russia, project
UR.01.01.287. We thank V. Yudin and D. Sarkisyan for helpful
discussion.

\end{document}